\documentclass[jkps,fleqn,showpacs,twocolumn,showkeys]{revtex4}
\usepackage{graphicx}
\usepackage{amssymb}
\usepackage{amsmath}
\usepackage{bm}
\usepackage{subfigure}
\begin{document}
\setcounter{page}{0}
\title{Boundary and finite-size effects in the competition between indirect magnetic exchange and Kondo screening}
\author{Irakli Titvinidze}
\email{irakli.titvinidze@physik.uni-hamburg.de}
\author{Michael Potthoff}
\affiliation{I. Institut f\"ur Theoretische Physik, Universit\"at Hamburg, Jungiusstra\ss{}e 9, 20355 Hamburg, Germany}

\begin{abstract}
A system of conduction electrons can mediate an indirect magnetic exchange between magnetic impurites. 
The nonlocal exchange typically competes with the local Kondo screening of the impurity magnetic moments.
In case of magnetic adatoms on non-magnetic surfaces this competition is expected to be affected by different confinement effects. 
Here we study this situation in the regime of intermediate hybridization strengths by means of real-space dynamical mean-field theory for a one-dimensional two-impurity Anderson model. 
Depending on the presence of a boundary and depending on the system size, a crossover between nonlocal and local singlet formation is observed in corresponding magnetic susceptibilities. 
The crossover is driven by the strong sensitivity of the Kondo temperature on the local density of states.
\end{abstract}

\pacs{71.10.Fd, 71.27.+a, 75.20.Hr, 75.75.-c} 

\keywords{Indirect magnetic exchange, Kondo effect, dynamical mean-field theory, magnetic adatoms}

\maketitle

{\em Introduction.}
Due to recent progress of experimental techniques to investigate magnetic adatoms on surfaces of non-magnetic metals not only the structural and electronic properties of adatoms and 
of the surrounding host material can be accessed, but also tailored magnetic model systems can be build \cite{Experiment}. One of the most fascinating effects in this context is the 
interplay between an indirect magnetic exchange of the adatoms via the substrate electrons, i.e.\ the Ruderman-Kittel-Kasuya-Yosida (RKKY) interaction \cite{RKKY}, on the one hand 
and Kondo screening \cite{Kondo} of the adatom magnetic moment by the conduction-band electrons of the substrate on the other.  
Impurities, step edges and different confinement effects may result in strong variations in the local substrate density of states. 
This strongly affects the Kondo screening as well as the RKKY interaction. 
Discrete eigenmodes can be found e.g.\ in quantum corrals located on a metallic surface \cite{MPE00}.
A quantum box, i.e.\ a quantum system confined in all three dimensions, represents the extreme case. 
Here, the local density of states is singular and consists of a set of delta-functions.
A quantum box can be realized, e.g., by semiconductor quantum dots \cite{CTL+04},
or as shortened carbon nanotubes \cite{OHCL00},
or as a single, e.g.\ metallocene, molecule \cite{BWD+05}.
Theoretically, the Kondo-box problem has been studied extensively \cite{paper_kondo_box,SA02,KZC+06,BBM+10}, while there is not so much is known on the competition between Kondo screening and RKKY interaction in a quantum box \cite{tunable_doniach}.

With the present contribution we report on studies of the effects of indirect magnetic exchange and its competition with Kondo screening for small one-dimensional model systems.
We will consider systems with open and periodic boundary conditions as well as thermodynamically large systems to disentangle boundary and finite-size effects.
Our main result is that there is a nontrivial competition between indirect magnetic exchange and Kondo screening which, at {\em intermediate} coupling strengths, is controlled by the local density of states rather than by finite-size gaps.

{\em Model.}
We consider the two-impurity Anderson model (TIAM):
\begin{eqnarray}
{\cal H}&=&-t\sum_{\langle i,j \rangle,\sigma}c_{i,\sigma}^\dagger c_{j,\sigma}^{\phantom\dagger}+U\sum_{\alpha=1}^{2}n_{\alpha,\uparrow}^f n_{\alpha,\downarrow}^f 
+\varepsilon\sum_{\alpha=1}^{2}n_{\alpha}^f  \\
&+&V \sum_{\alpha=1}^{2} \sum_\sigma\left(f_{\alpha,\sigma}^\dagger c_{i_{\alpha},\sigma}^{\phantom\dagger}+ h.c\right)-\mu\Bigl(\sum_{\alpha=1}^{2}n_{\alpha}^f+ 
\sum_{i=1}^{L} n_{i}^c \Bigl) \: . \nonumber 
\label{Hamiltonian}
\end{eqnarray}
Here $f_{\alpha,\sigma}^\dagger$ and $c_{i,\sigma}^\dagger$ are creation operators of an electron with spin projection $\sigma=\uparrow,\downarrow$ at the impurity (or ``adatom'') sites 
$\alpha=1,2$ or at the bath (``substrate'') sites $i=1,2,\ldots,L$, respectively. 
The corresponding occupation-number operators are $n_{\alpha,\sigma}^f= f_{\alpha,\sigma}^\dagger f_{\alpha,\sigma}^{\phantom\dagger}$ ($n_{\alpha}^f=n_{\alpha,\uparrow}^f+n_{\alpha,\downarrow}^f$) and 
$n_{i,\sigma}^c= c_{i,\sigma}^\dagger c_{i,\sigma}^{\phantom\dagger}$ ($n_{i}^c=n_{i,\uparrow}^c +n_{i,\downarrow}^c$). 
$t$ is the hopping amplitude between neighboring substrate lattice sites and is used to fix the energy scale, i.e.\ $t=1$. 
$U$ and $\varepsilon$ denote the on-site Hubbard interaction and the local on-site energy for the adatom sites. 
$\mu$ is chemical potential. $V$ is the hybridization between an adatom site $\alpha$ and the nearest-neighboring substrate lattice site which is denoted by $i_\alpha$. 
We consider the particle-hole symmetric model with $\mu=0$ and 
$\varepsilon = -U/2$ at half-filling which implies ground-state expectation values $\langle n_{\alpha}^f \rangle =1$ and $\langle n_{i}^c \rangle = 1$ for all $\alpha$ and $i$.

In the Kondo limit of the TIAM, where charge fluctuations on the adatom sites are strongly suppressed, there is an antiferromagnetic local exchange coupling $J\sim V^2/U$ between the adatom spin ${\bf S}_\alpha$ and the substrate spin ${\bf s}_{i_\alpha}$ by which one can characterize two relevant energy scales \cite{RKKY,Kondo}:
$J_{\rm RKKY}\propto J^{2}$ is 
the nonlocal indirect RKKY interaction and $T_{\rm K} \propto \exp(-1/|J|)$ the Kondo temperature. In the Kondo regime for $T_{\rm K} \gg |J_{\rm RKKY}|$ the local magnetic moments at 
the adatoms are independently screened by the substrate electron spins. 
On the other hand, in the RKKY limit for $T_{\rm K} \ll |J_{\rm RKKY}|$ and after integrating out the substrate degrees of freedom, the problem reduces to a two-spin Heisenberg model with exchange coupling 
$J_{\rm RKKY} = J^{2} \chi^{0,\rm sub}_{i_1 i_2}$, where $\chi^{0,\rm sub}_{i_1 i_2}$ is the nonlocal static susceptibility of the substrate. 

In the present study we will restrict ourselves to the case of intermediate hybridization strength $V$ to address a parameter regime where there are non-trivial effects of indirect magnetic exchange and which at the same time is clearly beyond the range of standard perturbative RKKY theory. 
To study systems of different size $L$, ranging from small quantum boxes of several tens of sites up 
to thermodynamically large systems where finite-size effects are absent, we employ the real-space dynamical mean-field theory (R-DMFT). 

{\em Method.}
R-DMFT \cite{RDMFT} is a comprehensive, thermodynamically consistent and non-perturbative mean-field approximation for correlated multi-impurity or lattice-fermion models with missing or reduced translational symmetry. 
The approach assumes the self-energy to be a local quantity but, in contrast to the standard DMFT \cite{DMFT}, retains its site-dependence. 
Each inequivalent and correlated site is self-consistently mapped onto a single-impurity Anderson model. 
The set of Anderson models can be solved independently to get the local but site-dependent self-energies. 
The different sites are coupled via Dyson's equation for the multi-impurity model the solution of which provides us with the local Green's functions.
These are used to determine the parameters of the impurity models in turn by means of the R-DMFT self-consistency equations. 
Details of the approach can be found in Refs.\ \cite{OUR,RDMFT}. 

In a preceding study \cite{OUR} for small one-dimensional systems ($L=50$) with open boundary conditions we have benchmarked the R-DMFT predictions against numerically exact density-matrix renormalization group (DMRG) calculations.
The mean-field approach turned out to be even quantitatively reliable in the case of intermediate and strong hybridization $V$. 

For small $V$, however, it was found that the R-DMFT breaks down:
Here the system becomes very susceptible to an externally applied magnetic field, and the mean-field approach artificially predicts a phase transition to a state with spontaneously broken symmetry.
Specifically, R-DMFT yields a mixture of degenerate states with ordered adatom local moments, $|\uparrow,\downarrow \rangle$ and $|\downarrow,\uparrow \rangle$, rather than the nonlocal singlet $(|\uparrow,\downarrow\rangle - |\downarrow,\uparrow\rangle ) / \sqrt{2}$ that is actually found in DMRG.

{\em Calculations.}
To study the TIAM at half-filling and zero temperature, the exact-diagonalization method \cite{ED} 
is used as an impurity solver.
The impurity Green's function is computed by means of the Lanczos technique for effective impurity models with $n_s=10$ sites.
We have checked that bath-discretization errors are negligible for the results discussed below.
To avoid a Kramers degeneracy of the ground state, we consider systems with an even number of substrate sites $L$. 
Calculations are done for one-dimensional systems with periodic boundary conditions. 
Small systems with $L=50$ and $L=52$ substrates sites as well as thermodynamically large systems are considered. 
$L=10000$ sites turned out to be sufficient to suppress finite-size effects in the calculation of the substrate Green's function (see Ref.\ \cite{OUR} for details).
To distinguish between finite-size and boundary effects we also compare the results with those of Ref.\ \cite{OUR} where open boundary conditions have been employed. 
In that case the two adatoms are coupled to substrate sites located symmetric to chain center.

{\em Observables.}
To study the magnetic properties of the system, we consider the local adatom susceptibilities $\chi_{11}=\chi_{22}$ and the nonlocal adatom-adatom susceptibilities $\chi_{12}=\chi_{21}$ as well as the adatom-substrate susceptibilities $\chi_{i\beta}^{\rm sub}$.
These are obtained as numerical derivatives of the respective local magnetic moment with respect to a local magnetic field $h_\beta$ coupling the $z$-component of the local adatom spin to the system via ${\cal H}\rightarrow{\cal H}-h_\beta S_{\beta,z}$: 
\begin{eqnarray}
\label{susceptibility}
\chi_{\alpha\beta}=\left.\frac{\partial m_\alpha^f}{\partial h_\beta}\right|_{h_\beta=0}\quad , \qquad
\chi_{i\beta}^{\rm sub}=\left.\frac{\partial m_i^c}{\partial h_\beta}\right|_{h_\beta=0} \: .
\end{eqnarray}
Here $m_\alpha^f=\langle {S_{\alpha,z}}\rangle=\frac{1}{2}(\langle n_{i,\uparrow}^f\rangle - \langle n_{i,\downarrow}^f \rangle)$ and  $m_i^c=\langle {s_{\alpha,z}}\rangle=
\frac{1}{2}(\langle n_{i,\uparrow}^c\rangle - \langle n_{i,\downarrow}^c \rangle)$ are magnetic moments on the adatom ($\alpha=1,2$) and on the substrate sites ($i=1,2,\ldots,L$), respectively.
 
\begin{figure}[t]
\includegraphics[width=0.45\textwidth]{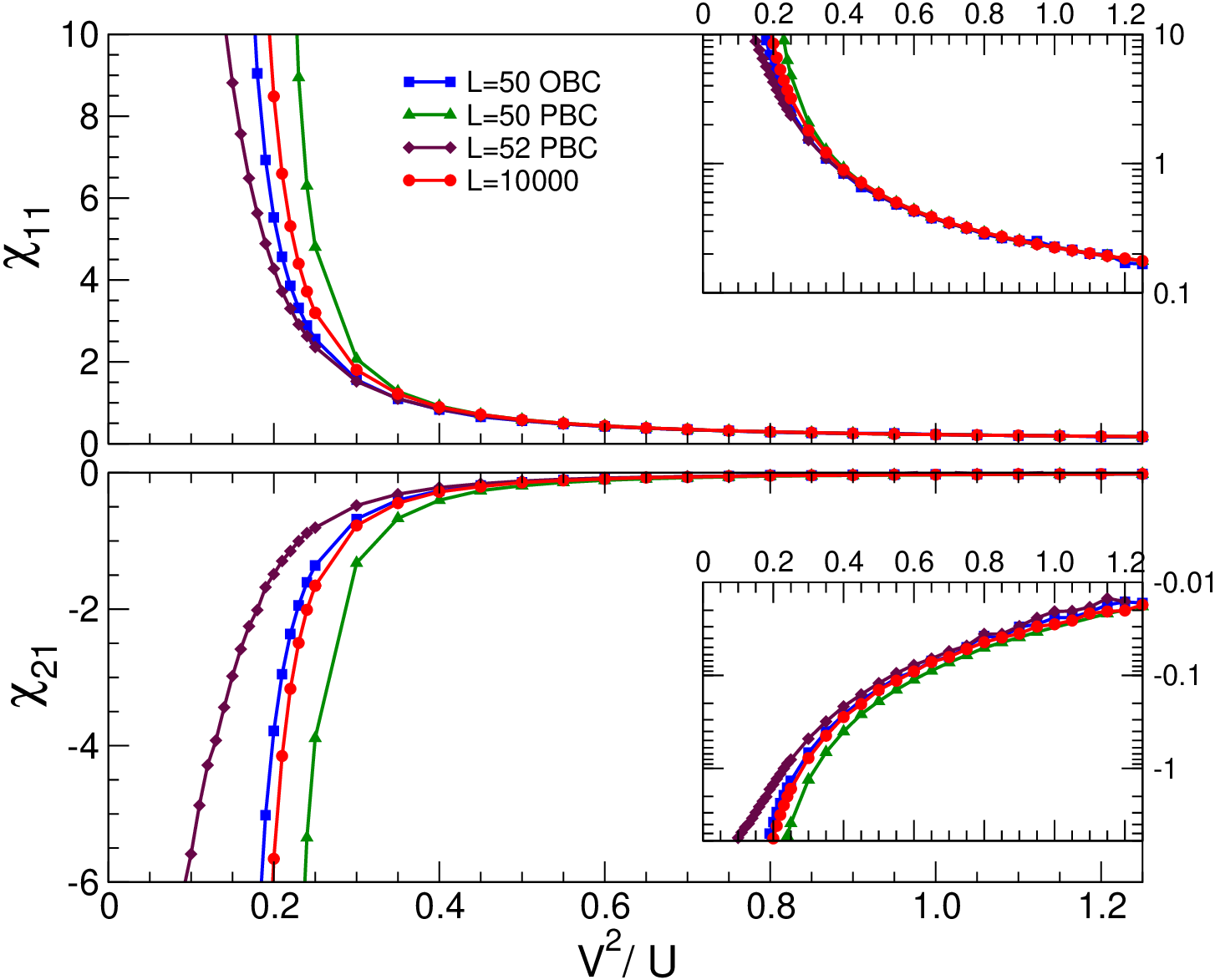}  
\caption{Local and nonlocal adatom susceptibilities $\chi_{11}$ and  $\chi_{21}$ as functions of $V^2/U$ for nearest-neighbor distance $d=1$ and $U=8$. Results are shown for different system sizes as indicated. 
The results for $L=50$ sites with open boundary condition are taken from Ref.\ \cite{OUR}.
Here the adatoms are located at positions symmetric to the chain center.
Insets: results plotted on a logarithmic scale.}
\label{Adatom-adatom_vs_J}
\end{figure}

\begin{figure*}[t]
\includegraphics[width=0.38\textwidth]{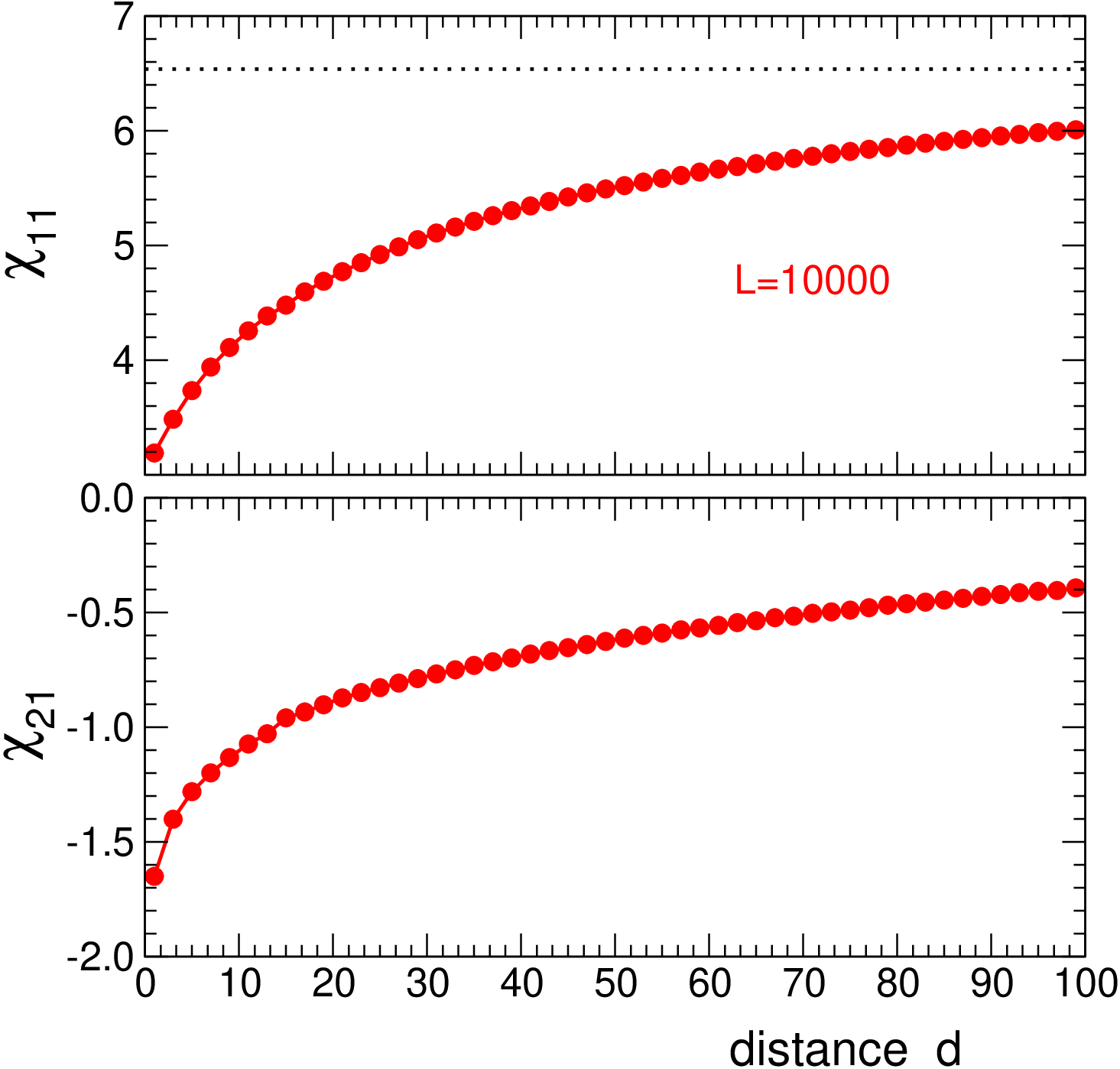}
\hspace{10mm}
\includegraphics[width=0.37\textwidth]{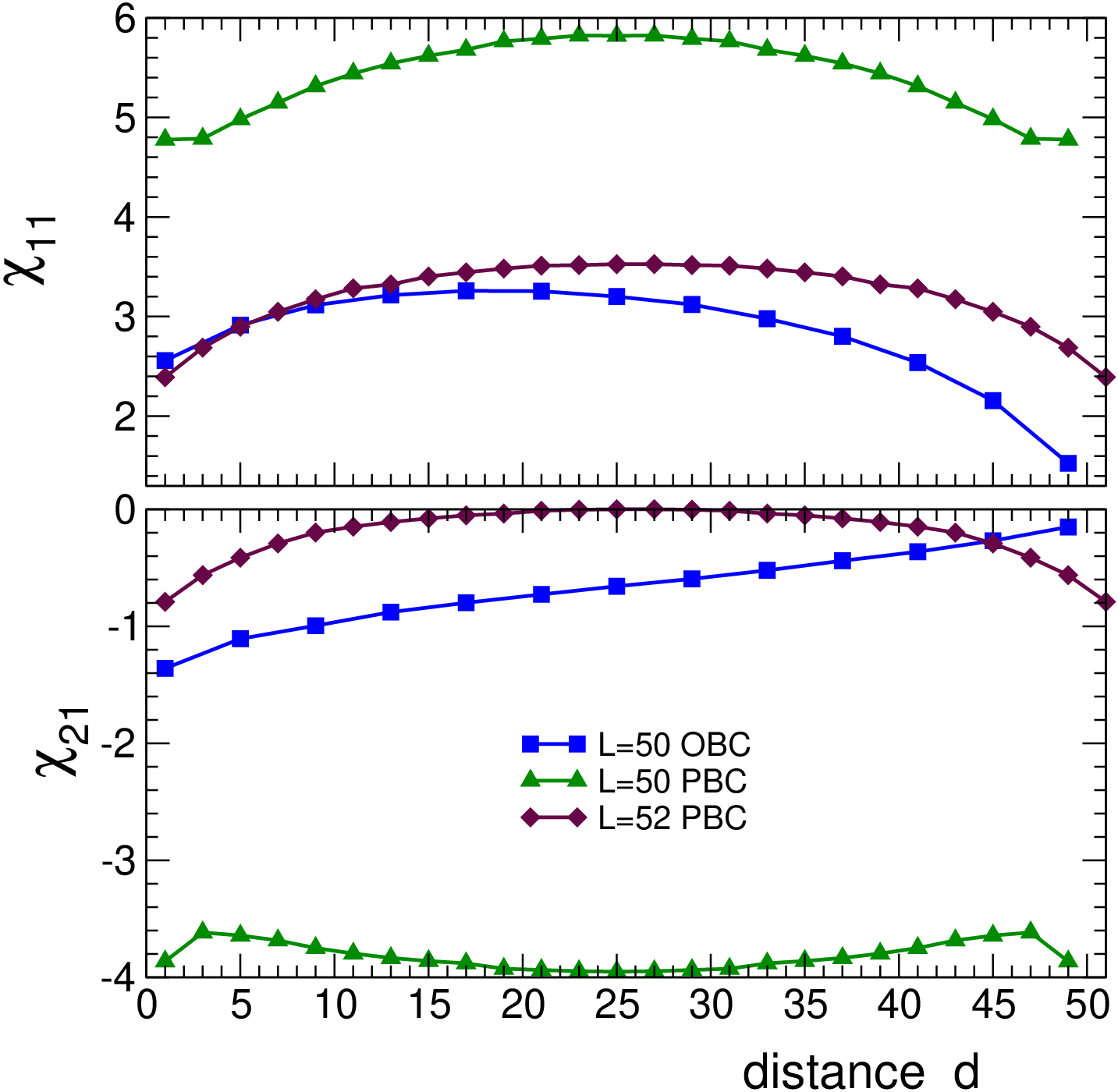}
\caption{Local and nonlocal adatom susceptibilities $\chi_{11}$ and  $\chi_{21}$ as functions of the distance $d$ between adatoms for $V^2/U=0.25$ and $U=8$.
Left: thermodynamically large system ($L=10000$). The dashed line indicates the single-impurity susceptibility $\chi_{11}^{\rm SI} = 6.54$.
Right: small systems ($L=50$ and $L=52$) for periodic and for open (taken from Ref.\ \cite{OUR}) boundary conditions.
}
\label{Adatom-adatom_vs_d}
\end{figure*}

{\em Dependence on the coupling strength.}
We first consider the case where the two adatoms are coupled to neighboring substrate sites, i.e.\ the distance between the adatoms is $d=1$ (see Fig.~\ref{Adatom-adatom_vs_J}). 
In the regime with antiferromagnetic Kondo coupling $V^2/U > t$, the susceptibilities for small systems ($L=50$ and $L=52$) are basically indistinguishable from those for a thermodynamically large system ($L=10000$). 
Contrary, for smaller values of $V^2/U$, the results start to depend on the size of the system and on the type of boundary conditions. 

This trend can be explained by referring to the concept of the Kondo cloud.
In the strong-coupling limit $V^2/U \gg t$, the size of the Kondo cloud $\xi$ is much smaller than size of the substrate $\xi \ll L$. 
Hence, finite-size effects in $\chi_{11}$, which roughly is proportional to $\xi$, are negligible. 
For distance $d=1$ boundary effects for open chains do not play a role either.

A further decrease of $V^2/U$ finally leads to a divergence of the susceptibilities (see Fig.~\ref{Adatom-adatom_vs_J}). 
This must be seen as an artifact of the R-DMFT and should be interpreted as an indication for the crossover to the perturbative RKKY regime. 
As Fig.\ \ref{Adatom-adatom_vs_J} shows, the position of this crossover is strongly affected by boundary and finite size effects. 

Here, we recall that besides the coupling strength, the Kondo temperature crucially depends on the local density of states. 
For a finite system, where the local density of states is composed of delta-peaks only, the relevant quantity is the weight $|U_{i_\alpha {k}_F}|^2$ of the highest occupied one-particle energy eigenstate $| k_F \rangle$ of the non-interacting substrate at $i_\alpha$, where $|k \rangle = L^{-1/2} \sum_i U_{ik} | i \rangle$ and $|i \rangle$ is the basis orbital at site $i$. The weight is site-dependent in case of open chains.
For $L=50$ one can easily see that $|U_{i_\alpha k_F}|^2$ is larger
compared to the weight for the infinite system.  
Consequently, 
the Kondo temperature is higher 
than for the infinite system. 
This explains, why the crossover to the RKKY regime takes place for smaller values of $V^2/U$ in case of open boundaries and $L=50$.

For $L=52$ sites and open boundaries (not shown in Fig.\ \ref{Adatom-adatom_vs_J}), we have just the opposite picture, and the Kondo temperature 
lower than for the infinite system. 
The difference between $L=4n$ and $L=4n+2$ (for integer $n$) results from scattering at the chain boundaries causing Friedel oscillations in the weight with a wave vector $2 k_F = \pi$. 

Contrary, for systems with periodic boundaries, the weight $|U_{i_\alpha {k}_F}|^2=1$ is site-independent and also independent of the system size.
Still there is a clear oscillation visible in the susceptibilities between systems with 
$L=4n$ and $L=4n+2$ sites.
Here, the decisive difference is that there is a (twofold degenerate) eigenenergy at zero excitation energy for $L=4n$ and thus a stronger effective coupling of the adatom moment and a higher $T_{\rm K}$ while there is a gap in the local density of states for $L=4n+2$.

We conclude that, depending on the hybridization strength, finite-size and boundary effects show up in local adatom susceptibility $\chi_{11}$ which can be explained by a competition between the Kondo effect and indirect magnetic exchange.
As a divergency of $\chi_{11}$ also implies $\chi_{21}$ to be divergent, the same trends are also found there.
Anyway, in the RKKY limit $J\to 0$, we have the sum rule $\chi_{11} + \chi_{21}=0$, and the behavior of $\chi_{21}$ (see Fig.\ \ref{Adatom-adatom_vs_J}) can also be interpreted as being a reminiscence to the RKKY limit.

{\em Infinite system: distance dependence.}
Fig.~\ref{Adatom-adatom_vs_d} (left) displays our results for the local and the nonlocal adatom-adatom susceptibilities as a function of the distance $d$ for fixed $V^2/U=0.25$.
At this coupling strength, $L=10000$ substrate sites are sufficient to simulate the infinite system.
Here, and in the following, we always consider odd distances $d$ which results in an antiferromagnetic indirect exchange.
The ground state is a spin singlet. 

The modulus of $\chi_{21}$ decreases with increasing distance $d$, contrary to what would be predicted by an effective RKKY two-spin model since a decreasing coupling ($J_{\rm RKKY} \propto 1/d$ in one dimension at half-filling) would lead to more susceptible moments.
Here, rather a picture of two separate Kondo clouds applies where with increasing $d$ the adatom-substrate coupling becomes more important.
Recall that for a gapped system we trivially have the sum rule $\chi_{11} + \chi_{21} + \sum_i\chi_{i1}^{\rm  sub}=0$. 
The large-$d$ limit is equivalent with the single-impurity (SI) limit. 
This implies $\chi_{11} \to \chi_{11}^{\rm SI}$ and
$\sum_i \chi_{i1}^{\rm sub} \to \sum_i \chi_{i1}^{\rm sub, SI}$ with 
$\chi_{11}^{\rm SI} + \sum_i \chi_{i1}^{\rm sub, SI} = 0$ and therefore 
$\chi_{21} \to 0$.

The local adatom susceptibility $\chi_{11}$, on the other hand, increases with $d$ and very slowly converges to $\chi_{11}^{\rm SI}$. 
We find $\chi_{11}^{\rm SI}=6.54$ (see dashed line in Fig.~\ref{Adatom-adatom_vs_d} (left)) from a non-selfconsistent calculation for a single adatom.

{\em Finite systems: boundary effects.}
Fig.~\ref{Adatom-adatom_vs_d} (right) shows corresponding results for small systems.
We first discuss the results obtained for open boundaries (see blue lines). 
While the nonlocal adatom-adatom susceptibility $\chi_{21}$ again tends to zero with increasing $d$, we find a non-monotonic behavior of the local adatom susceptibility $\chi_{11}$.
As detailed in Ref.\ \cite{OUR}, the decrease of $\chi_{11}$ for large $d$ is a boundary effect. 
With increasing $d$ the adatoms move to the chain edges. 
This implies an increase of the local substrate density of states at the adatom sites which gives rise to an increase of the local Kondo temperature and thus a decrease of the local susceptibility. 

\begin{figure}[t]
\includegraphics[width=0.4\textwidth]{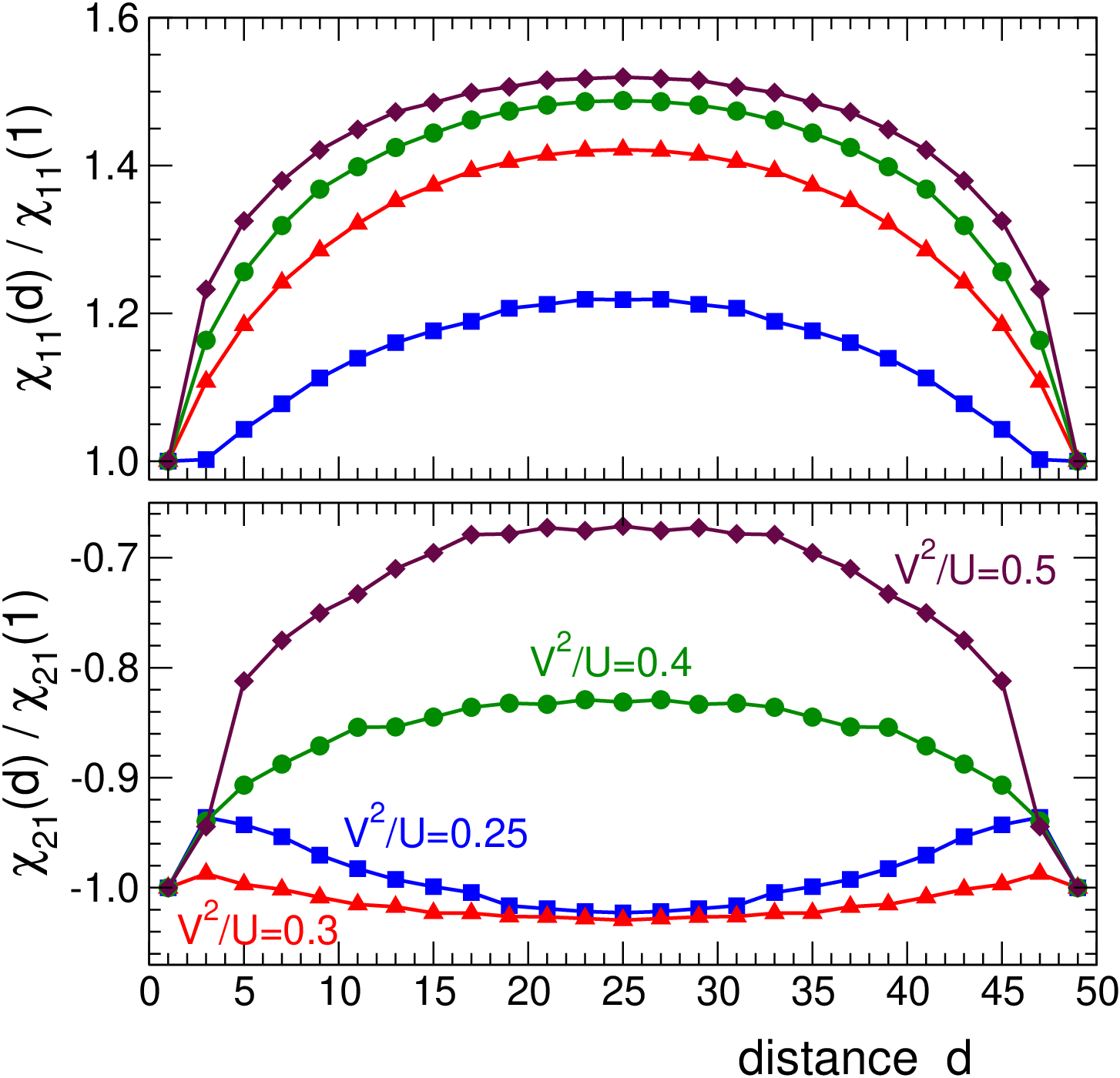}
\caption{
$\chi_{11}$ and  $\chi_{21}$ as functions of $d$ for $U=8$ and $L=50$ using periodic boundary conditions.
Results for different $V^2/U$ as indicated. 
Note that the susceptibilities are rescaled with their value for $d=1$.
}
\label{chipbc}
\end{figure}

{\em Finite systems with periodic boundaries.}
For periodic boundary conditions, the situation should be more simple as edge effects are absent. 
The results obtained for systems with $L=52$ substrate sites are indeed qualitatively similar to those for the infinite system (Fig.~\ref{Adatom-adatom_vs_d} (right), see brownish lines).
Note that the maxima seen in $\chi_{11}$ and $\chi_{21}$ at $d=(L+1)/2$ simply result from the equivalence between distances $d$ and $L-d$. 

For $L=50$, on the other hand, the nonlocal susceptibility $\chi_{21}$  behaves quite different. 
Its modulus $|\chi_{21}|$ {\em increases} with increasing distance for $d<(L+1)/2$ (except for the extreme case of nearest neighbors, i.e.\ $|\chi_{21}(d=3)|<|\chi_{21}(d=1)|$).
An increase of $|\chi_{21}|$ with increasing distance between the adatoms could be understood perturbatively within RKKY theory as mentioned above. 
However, $V^2/U=0.25$ and $U=8$ would refer to the strong-coupling limit ($J=2$ in units of $t$) and, anyway, there are still significant charge fluctuations at the adatom sites, and hence the RKKY picture does not apply. 
Nevertheless, it is tempting to interpret the trend of $|\chi_{21}|$ as a reminiscence to the perturbative RKKY regime. 
Namely, as discussed above, the system has a lower Kondo temperature compared to the infinite system and compared to the case $L=52$, too.
That the formation of a nonlocal singlet is favored here, is also corroborated by the larger absolute values of 
$\chi_{21}$ and by the fact that $\chi_{21}$ behaves almost complementary to $\chi_{11}$ as it is characteristic for the RKKY limit where $\chi_{11}  + \chi_{21} = 0$ holds.

\begin{figure}[t]
\includegraphics[width=0.4\textwidth]{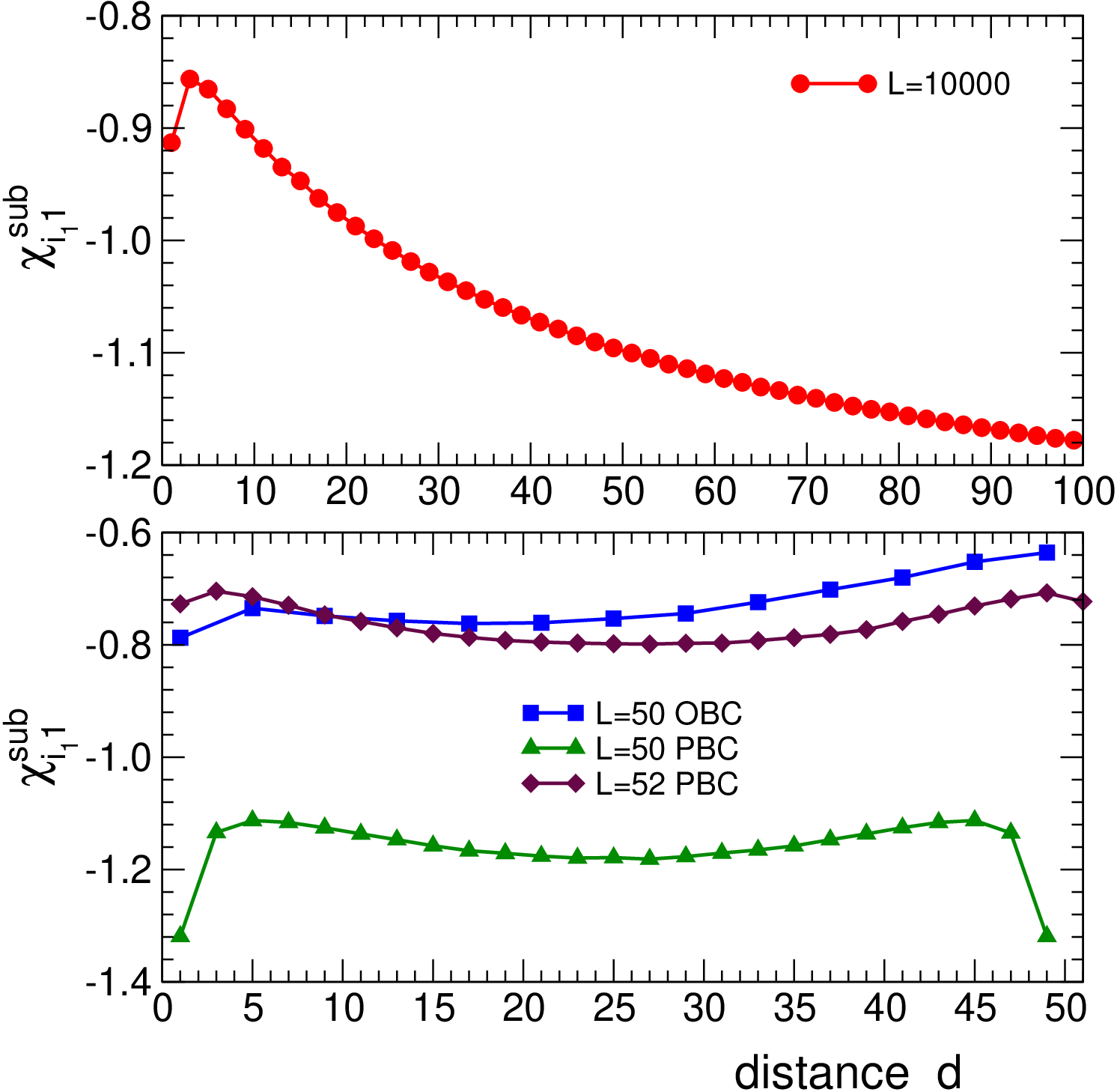}  
\caption{Local adatom-substrate susceptibility $\chi_{i_11}$, where $i_1$ is the site ``below'' the first adatom, as a function of the distance $d$ to the second adatom. 
Results for $V^2/U=0.25$ and $U=8$.
Upper panel: large system, $L=10000$.
Lower panel: small systems, $L=50$ and $L=52$, with periodic and open (taken from Ref.\ \cite{OUR}) boundary conditions.
}
\label{local_substrate_chi_vs_d}
\end{figure}

{\em Crossover between local and nonlocal singlets in finite systems.}
As a further check we have performed calculations for $L=50$ using periodic boundary conditions again but at larger values of $V^2/U$ (see Fig.\ \ref{chipbc} and note that the results are rescaled for better comparison).
In fact, with increasing $V^2/U$ the nonlocal adatom-adatom susceptibility $\chi_{21}$ nicely shows a crossover from the regime where nonlocal singlet formation dominates to a Kondo regime with local singlets.
This is indicated by the qualitative and systematic change of the $d$ dependence of $\chi_{21}$ when increasing $V^2/U$.

With this we can also understand the anomalous behavior of the susceptibility at $d=1$ compared to $d=3$:
Due to the proximity to the RKKY regime, the feedback of nonlocal magnetic correlations on the selfenergy is increased. 
Since this is neglected within R-DMFT, the behavior is most probably an artifact of the mean-field approximation. 
Consistent with this are the slight deviations of the R-DMFT results from numerically exact DMRG data that have been observed in Ref.\ \cite{OUR} at $d=1$ for similar model parameters.

Finally, Fig.\ \ref{local_substrate_chi_vs_d} displays the results for the local adatom-substrate susceptibility $\chi^{\rm sub}_{i_11}$ which measures the response of the substrate site $i_1$ ``below'' the adatom to a local field at the same adatom.
This probes the strength of local Kondo singlet formation.
$\chi^{\rm sub}_{i_11}$ in fact is negative in all cases. 
Consistent with the above interpretation the modulus $|\chi^{\rm sub}_{i_11}|$ is the largest for the system with the lowest $T_{\rm K}$, i.e.\ for $L=50$ with periodic boundaries.
For open boundaries, we again have a site-dependent Kondo temperature. 
This explains the non-monotonic behavior of $|\chi^{\rm sub}_{i_11}|$ as a function of the distance to the second adatom
(note that for periodic boundaries we trivially have an equivalence between distances $d$ and $L-d$). 
We also find the artifacts of the mean-field approximation to be the strongest for systems with dominant nonlocal singlet formation, i.e.\ for $L=50$ with periodic boundaries.

{\em Conclusions.} 
We have studied model systems where two correlated sites with well-defined local magnetic moments interact locally with the itinerant moments of non-interacting conduction electrons.
As has been verified previously by comparison with DMRG calculations, a regime of intermediate hybridization strengths is well accessible to real-space dynamical mean-field theory.
Here we find that just in this parameter range a non-trivial competition between nonlocal indirect exchange and local Kondo screening takes place which is strongly affected by finite-size and boundary effects.
This is most clearly seen in the static nonlocal inter-impurity magnetic susceptibility. 
Depending on the type of boundary conditions and depending on the system size, a crossover between different distance dependencies that are characteristic for the two extreme cases of the nonlocal RKKY singlet and local Kondo singlets respectively, is found as a function of the coupling strength.
Rather than a low-energy cutoff due to finite-size gaps, the main mechanism controlling this competition at intermediate hybridization strength is the exponentially strong dependence of the Kondo temperature on the local density of states. 
The LDOS in turn is sensitively affected by the system size and according odd-even effects and, via Friedel oscillations, by the presence of the system's boundary.
While a one-dimensional model system has been the focus of the present study, real-space DMFT can easily be applied to basically arbitrary structures of magnetic adatoms on surfaces of higher-dimensional lattices. 
Corresponding studies are intended for the future.

\begin{acknowledgments}
We would like to thank A.\ Schwabe for instructive discussions.
The work has been supported by the Deutsche Forschungsgemeinschaft within the Sonderforschungsbereich 668, project A14.
\end{acknowledgments}

\end{document}